\documentclass[letterpaper,journal]{IEEEtran}
\usepackage{amsmath,amsfonts}
\usepackage{algorithmic}
\usepackage{algorithm}
\usepackage{array}
\usepackage[caption=false,font=footnotesize,labelfont=sf,textfont=sf]{subfig}
\usepackage{textcomp}
\usepackage{stfloats}
\usepackage{url}
\usepackage{verbatim}
\usepackage{graphicx}
\usepackage{cite}
\usepackage[table,xcdraw]{xcolor}

\begin{document}

\title{Evaluating Architectural Trade-offs in CGRAs: The Impact of Scratchpad Memory and Heterogeneity on Compute-Intensive Kernels}

\author{María José Belda, Lara Orlandic, Fernando Castro, Miguel Peón-Quirós, Katzalin Olcoz, David Atienza% <-this % stops a space
\thanks{M. J. Belda, F. Castro, and K. Olcoz are with Complutense University of Madrid, Madrid, Spain (Correspondence author: mbelda@ucm.es). L. Orlandic and D. Atienza are with Embedded Systems Laboratory (ESL), EPFL, Lausanne, Switzerland. M. Peón-Quirós is with EcoCloud, EPFL, Lausanne, Switzerland.}%
\thanks{This paper has been partially funded by the EU (FEDER), the Spanish MINECO under grants PID2021-126576NB-I00 and PID2024-158311NB-I00 funded by MCIN/AEI/10.13039/501100011033 and by European Union “ERDF A way of making Europe” and the NextGenerationEU/PRT.}% <-this % stops a space
\thanks{Manuscript received Month DD, YYYY; revised Month DD, YYYY.}%
}

\markboth{IEEE Transactions on Computer-Aided Design of Integrated Circuits \& Systems,~Vol.~XX, No.~X, Month~202X}%
{Belda \MakeLowercase{\textit{et al.}}}


\maketitle

\begin{abstract} % 150-200 words
Modern edge computing applications, particularly high-throughput stream processing like Vision Transformers (ViTs), demand massive spatial parallelism and efficient data movement under tight power and area constraints. Coarse-Grained Reconfigurable Architectures (CGRAs) offer a promising paradigm to balance performance, flexibility, and energy efficiency. This paper analyzes the impact of two critical CGRA design choices: processing element heterogeneity and local data reuse support. We evaluate essential computational kernels—Fast Fourier Transform (FFT) and General Matrix Multiply (GEMM)—alongside an end-to-end seizure detection transformer workload across two distinct configurations: a baseline homogeneous architecture and a heterogeneous evolution integrating specialized functional units with an Scratchpad Memory (SPM). Our evaluation demonstrates that the SPM significantly optimizes data movement, reducing memory traffic eightfold compared to a memory-less design. While the heterogeneous architecture achieves superior energy efficiency for data-shuffling tasks, the homogeneous design minimizes area overhead by 4.4$\times$ to 8.2$\times$ relative to state-of-the-art CGRAs. Furthermore, it sustains a 700 MHz operating frequency, enabling up to a $5\times$ execution speedup over the heterogeneous configuration during matrix computations. Ultimately, this work provides an architectural roadmap for selecting CGRA fabrics based on the arithmetic intensity, performance goals, and resource envelopes of edge-scale workloads.
\end{abstract}

\begin{IEEEkeywords} % 4-8 Keywords
CGRA, Heterogeneous Computing, SPM, Data reuse optimization, FFT, GEMM, Hardware-software co-design
\end{IEEEkeywords}

% ------------------------------------------------
%               I. INTRODUCTION
% ------------------------------------------------
\section{Introduction}

\IEEEPARstart{T}{he} rapid evolution of data-intensive computing has shifted the focus of edge processing toward high-performance execution of complex algorithms under extreme energy and area constraints. Modern edge workloads are increasingly dominated by stream-oriented, multi-dimensional data kernels that require massive spatial parallelism and optimized data movement to overcome the ``memory wall" inherent to resource-constrained environments~\cite{ma2023tsd, najafi2024versasens, khan2022transformers, prabhakar2017plasticine}. Achieving tight architectural efficiency is especially critical for real-time sensing applications in wearable and embedded devices, where the balance between low-latency performance and ultra-low-power (ULP) operation is paramount~\cite{sanchez2025low}. Although Application-Specific Integrated Circuits (ASICs) offer maximum efficiency for fixed dataflow graphs, they lack the flexibility to adapt to changing algorithmic requirements. In contrast, FPGAs provide reconfigurability but incur a significant ``energy tax" due to their bit-level routing making them less ideal for ULP field deployment~\cite{calhoun2010flexible, kuon2008fpga, poon2005detailed}.

CGRAs have emerged as a leading architectural paradigm to bridge this gap, reconciling ASIC-like computational efficiency with software-like programmability~\cite{liu2019survey, podobas2020survey, bandara2022revamp}. By operating at word-level granularity, CGRAs natively execute vector operations and match the datapath width of streaming algorithms, drastically reducing control logic and interconnect overhead compared to fine-grained fabrics~\cite{sutter2018coarse, aliagha2022energy}. This architectural efficiency is critical for edge-scale deployment, where off-chip communication is orders of magnitude more expensive than local computing, necessitating hardware that maximizes data reuse and local storage through specialized memory hierarchies~\cite{kim2025hastily}.

In this work, we explore two critical CGRA design dimensions: (i) the trade-off between homogeneous fabrics and heterogeneous specialized units, and (ii) the efficiency of direct memory access (DMA) versus software-managed scratchpad-based data reuse. While homogeneous CGRAs simplify execution mapping for regular, symmetric dataflow graphs (DFGs), heterogeneous designs enable the integration of specialized acceleration units—such as dedicated arithmetic slots, complex loop-control units, or SIMD-capable functional blocks—that can significantly amortize the energy cost of repetitive mathematical operations. 

However, navigating processing elements (PE) heterogeneity during compilation introduces substantial mapping and orchestration overheads, particularly when distinguishing between specialized load/store units (LSUs) and compute-only functional PEs~\cite{kou2022taem, dai2025dependency}. Managing this structural variation often requires advanced control specialization frameworks, where dedicated hardware controllers or specialized I/O blocks are deployed to handle loop control and data synchronization across multitasking execution environments~\cite{yang2025multisky, dai2025dependency}. Furthermore, optimizing data staging within these heterogeneous configurations necessitates strict coordination across localized register hierarchies, balancing local register files, global register files, and delay pipes to prevent interconnect routing congestion~\cite{li2026transmap, kou2022taem}. Beyond computing element orchestration, spatial CGRA performance in edge-scale workloads is heavily throttled by physical memory constraints. To prevent memory bottlenecks from stalling the execution array, limited SPM structures must be optimized to match the workload's data parallelism and maintain high PE utilization~\cite{chen2025data, dai2025dependency}.

This paper presents a comparative evaluation of two architectures tailored for high-throughput edge workloads: OpenEdgeCGRA~\cite{alvarez2023open}, a homogeneous array, and the DomaIn-specific System-technology CO-design (DISCO) CGRA, a heterogeneous architecture featuring specialized PEs and a shared SPM. We use the baseline kernels of FFT and GEMM, alongside an end-to-end Transformer-based seizure detection application (TSD), as a case study to evaluate execution time, power, and memory efficiency across both architectures.

The main contributions of this work are:
\begin{itemize}
    \item We quantify the performance, power dissipation, and energy consumption trade-offs between homogeneous and heterogeneous CGRA designs, identifying how architectural specialization accelerates data-intensive applications.
    \item We establish architectural guidelines for selecting CGRA configurations based on the application domain, focusing on the trade-offs between local SPM availability and data reuse potential in embedded matrix and tensor operations.
    \item We enhance DISCO base architecture with specialized Multiply-Accumulate (MAC) and Single Instruction, Multiple Data (SIMD) units, optimizing the design for high-throughput, energy-constrained streaming acceleration.
    \item We develop a manually optimized GEMM implementation for the DISCO platform to enable high-performance execution of Transformer-based attention layers in the absence of automated compiler toolchains.
\end{itemize}

Ultimately, our comparative evaluation provides a roadmap for selecting between homogeneous and heterogeneous fabrics based on the target deployment domain—ranging from ultra-low-power implantable electronics to throughput-driven or autonomous monitoring systems.

The remainder of this paper is organized as follows. Section~\ref{sec:background} reviews the background on CGRA fabrics and memory hierarchies. Section~\ref{sec:architectures} presents the technical specifications of the homogeneous OpenEdgeCGRA and heterogeneous DISCO-CGRA target architectures. Section~\ref{sec:exp_setup} discusses kernel mapping strategies, followed by the introduction of an end-to-end seizure detection transformer case study and details the measurement setup. Section~\ref{sec:results} analyzes the performance, power, and energy results. Finally, Section~\ref{sec:conclusions} concludes the paper with a summary of architectural trade-offs and selection guidelines.

% ------------------------------------------------
%                  II. BACKGROUND
% ------------------------------------------------
\section{Background and Architectural Overview}
\label{sec:background}

CGRAs have been extensively studied as an efficient alternative to both general-purpose processors and fine-grained reconfigurable fabrics, achieving a favourable balance between performance, energy efficiency, and programmability by exploiting spatial parallelism and word-level reconfiguration~\cite{liu2019survey, podobas2020survey}. Although existing surveys provide broad classifications of CGRA architectures, they do not directly analyse how architectural aspects such as PE heterogeneity and memory hierarchy organization influence the target application domain and other design metrics, including energy consumption and area. This work addresses this gap through a comparative study of these architectural features and their impact on CGRA behaviour and efficiency.

\subsection{Functional Heterogeneity in CGRA Fabrics}

Traditional homogeneous CGRAs, such as MorphoSys~\cite{singh2000morphosys} and OpenEdgeCGRA~\cite{alvarez2023open}, utilize regular arrays of identical reconfigurable cells. This approach is also adopted by the REVAMP~\cite{bandara2022revamp} framework's canonical baselines, including HM-ADRES and HM-HyCUBE, where every PE contains a full-fledged ALU. While these designs simplify hardware and mapping by offering uniform resources, they often lead to inefficiencies and underutilization when workloads do not fully exploit every PE's capabilities. 

To address these limitations, heterogeneous CGRAs incorporate specialized functional units (FUs) to optimize area and energy for specific tasks. For instance, RipTide~\cite{gobieski_riptide_2022} utilizes PEs specialized for arithmetic and memory, while its programmable NoC routers serve as specialized units for control operations. Plasticine~\cite{prabhakar2017plasticine} distinguishes between Pattern Compute Units (PCUs) for SIMD processing and Pattern Memory Units (PMUs) for banking and addressing logic. Similarly, SNAFU~\cite{gobieski_snafu_2021} employs a ``bring your own functional unit'' (BYOFU) approach with a standard interface to integrate diverse units like multiplication or global memory access. The VWR2A~\cite{denkinger2022vwr2a} architecture adopts a hybrid organization where a core array of reconfigurable cells is augmented with specialized slots, such as Loop-Control Units (LCU) for branch management and Load-Store Units (LSU) for high-bandwidth data movement, effectively offloading control-intensive tasks from the general-purpose fabric. Other designs, such as ULP-SRP~\cite{kim2014ulp}, restrict complex instructions like division to a few specialized units to save power, while R-Blocks~\cite{de_bruin_r-blocks_2024} employs five distinct FU types, such as the Branch Unit (BU), to reuse instruction encoding space and increase flexibility. Templates like ADRES~\cite{mei2003adres} support heterogeneous operation sets, and the REVAMP~\cite{bandara2022revamp} framework extends this by generating variants like HT-ADRES and HT-HyCUBE, which distribute operations based on application frequency to reduce area overhead.

\subsection{Memory Hierarchy and Data Locality}

Efficient data management is further achieved through specialized memory hierarchies, most notably SPMs, which improve energy efficiency by moving control to the software side. The VWR2A~\cite{denkinger2022vwr2a} architecture specifically integrates a 32~KiB shared SPM alongside very-wide registers (VWRs) to offer performance similar to caches but at a lower energy cost. In Plasticine~\cite{prabhakar2017plasticine}, the PMUs contain banked scratchpads that can be configured for stride, FIFO, or N-buffering access patterns. Similarly, SNAFU~\cite{gobieski_snafu_2021} and RipTide~\cite{gobieski_riptide_2022} utilize 1~KiB SRAM scratchpad PEs to hold intermediate data communicated between fabric configurations when a dataflow graph is too large to fit at once. 

R-Blocks~\cite{de_bruin_r-blocks_2024} features multiple 1~KiB local memory units with two-port arbiters and reconfigurable state registers for parallel access. Other strategies include the MorphoSys~\cite{singh2000morphosys} frame buffer, which is logically organized into two sets to overlap computation with DMA transfers, and the ULP-SRP~\cite{kim2014ulp} architecture, which features a Unified Memory Architecture (UMA) that allows users to partition a 512~KiB block between instruction and data regions. Finally, the REVAMP~\cite{bandara2022revamp} framework introduces heterogeneity into PE-local storage, deriving variants where memory sizes are non-uniform across the array, allocating more resources to ``hotspot'' PEs. While these specialized memory and heterogeneous structures significantly enhance efficiency, they inherently increase the area and the complexity of the compiler's mapping task.

% ------------------------------------------------
%               III. TARGET ARCHITECTURES
% ------------------------------------------------
\section{Target Architectures}
\label{sec:architectures}
The selection of OpenEdgeCGRA (OE) and DISCO-CGRA (DISCO) for this study is based on their shared functional \mbox{heritage}. As DISCO represents an architectural evolution of the OE model, and both fabrics support an equivalent set of operations. This relationship enables a controlled comparison where the impact of specific architectural features—namely structural heterogeneity and the integration of a local scratchpad—can be isolated and evaluated using identical computational kernels adapted for both environments.

\subsection{OpenEdgeCGRA: Homogeneous and Memory-Less}
\begin{figure}[t]
    \centering
    \includegraphics[width=0.9\linewidth]{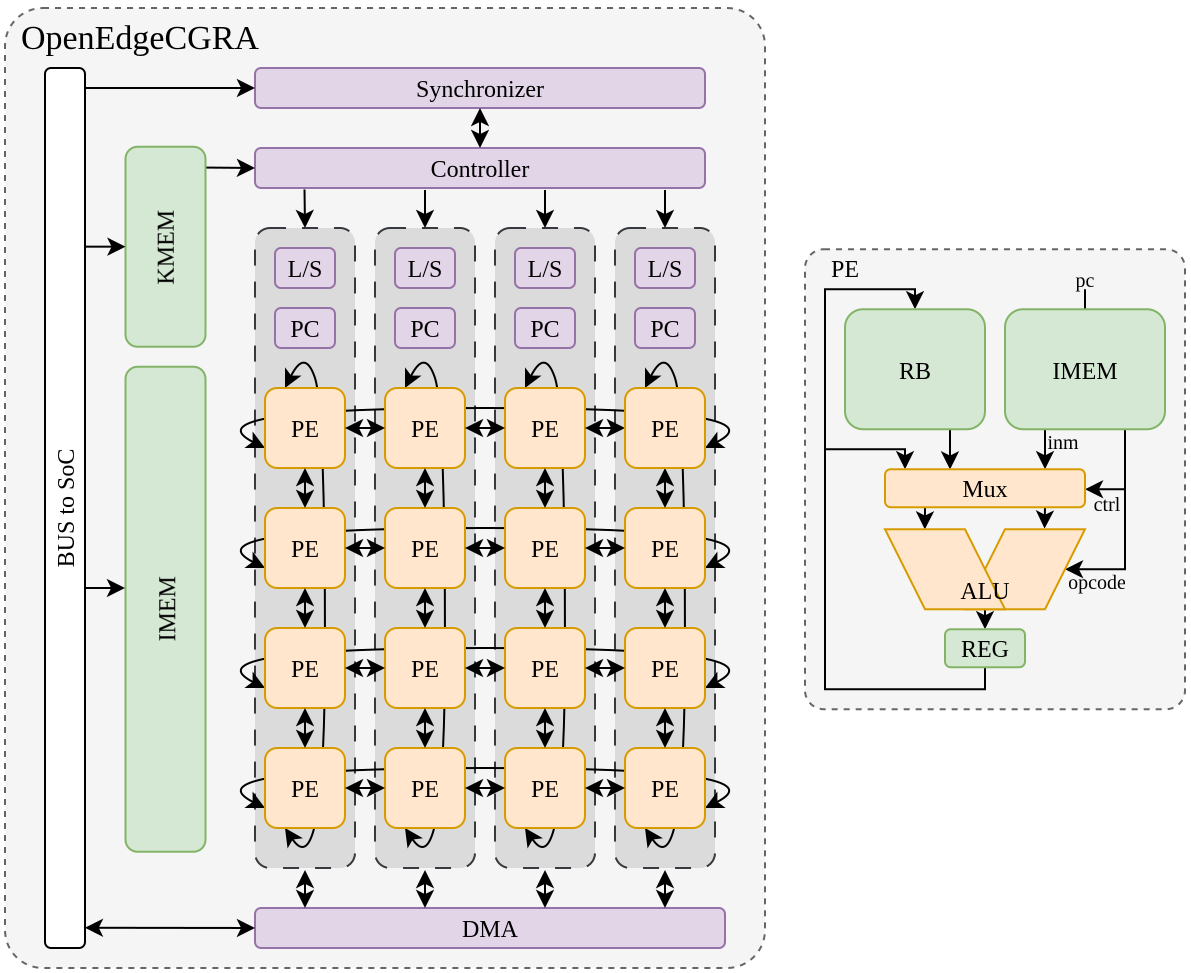}
    \caption{OpenEdgeCGRA architecture scheme. Green, purple, and orange denote memory, control and computation, respectively.} 
    \label{fig:oe_arch}
\end{figure}

OE is a time-distributed, homogeneous architecture employing a multiple-configuration, multiple-data model with static scheduling. As illustrated in Figure~\ref{fig:oe_arch}, the fabric is organized as a $4\times4$ grid of identical PEs connected via a torus topology. 

The architecture is characterized by its memory-less data plane, lacking a local scratchpad and relying on a DMA-backed interface to main memory for all operand movements. Control is managed at the column level, with each column possessing an independent program counter (PC) and DMA connection. Instructions are 32 bits wide and follow a hierarchical storage model: for each kernel, instructions are held in a global instruction memory (IMEM) and loaded into the PEs' local IMEMs upon execution. Additionally, a specific kernel IMEM (KMEM) stores the necessary configuration for each kernel. At the processing level, each PE shares the same internal structure, featuring an ALU connected to a local Register Bank (RB).

\subsection{DISCO-CGRA: Heterogeneous with Local Scratchpad}
\begin{figure}[t]
    \centering
    \includegraphics[width=0.825\linewidth]{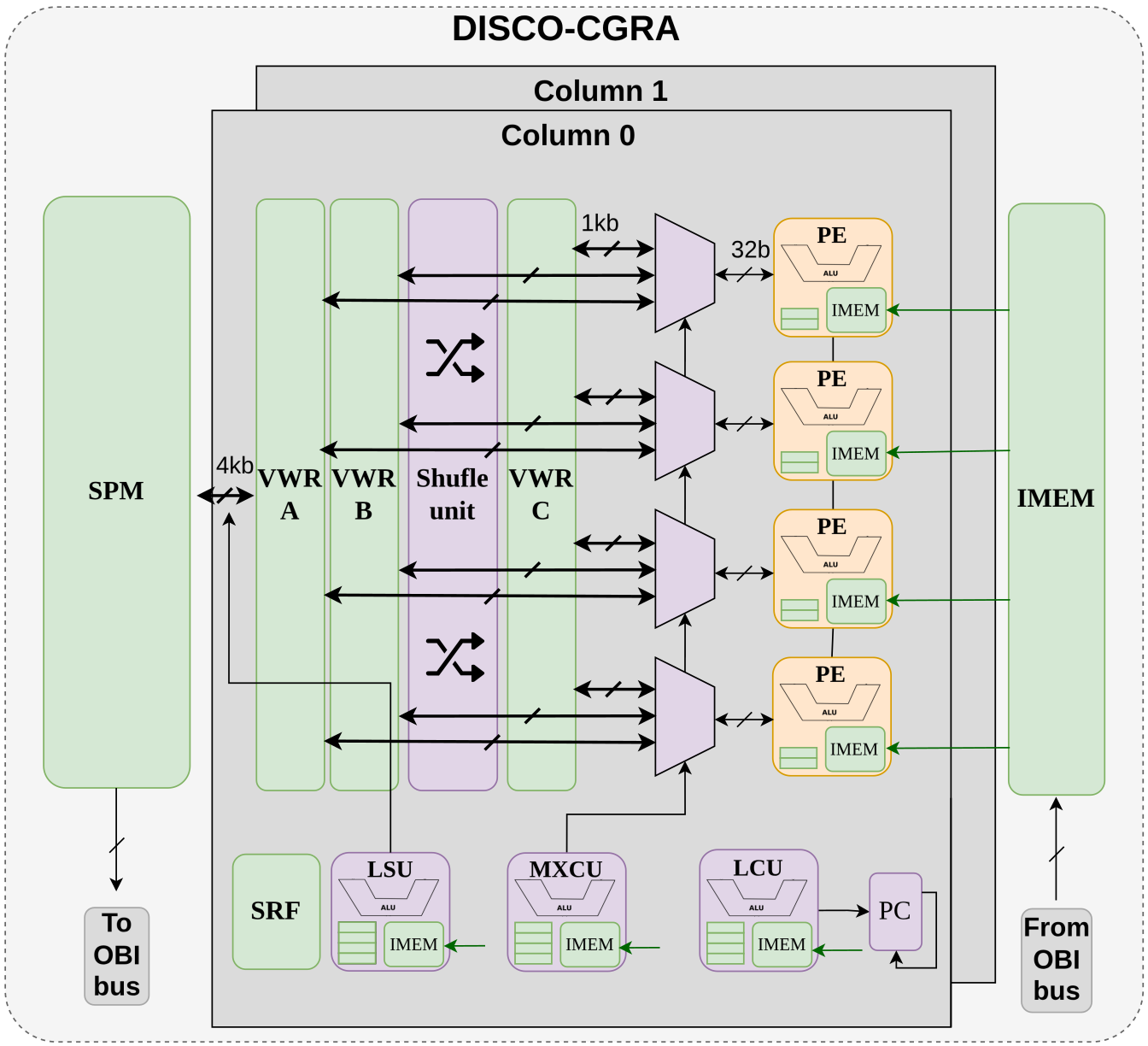}
    \caption{DISCO-CGRA architecture scheme. Green, purple, and orange denote memory, control and computation, respectively.} 
    \label{fig:disco_arch}
\end{figure}

DISCO represents an optimized architecture based on VWR2A~\cite{denkinger2022vwr2a}, which retains the core computational capabilities of the OE design. While VWR2A introduced a unique interconnect utilizing very wide memory elements, DISCO incorporates specific architectural optimizations—specifically SIMD capabilities and MAC support on the PEs—to enhance performance and energy efficiency. As illustrated in Figure~\ref{fig:disco_arch}, the architecture is structured into two independent columns. Each column operates as a self-contained processing cluster, integrating four PEs, three VWRs, a local PC, and a Scalar Register File (SRF) with shared access for all functional units. Within this structure, heterogeneity is achieved through three specialized hardware units: the LSU, LCU, and Multiplexer-Control Unit (MXCU).

The memory hierarchy is centered around an SPM interfaced with the VWRs. In this work, each VWR features a configurable width of 4096 bits and can be loaded from the SPM in a single clock cycle. Access to these registers is distributed across the column, with each PE capable of accessing a 1024-bit segment (one-fourth) of each VWR.

The control plane follows a VLIW-style execution model where seven simultaneous, distinct instructions are executed per column every clock cycle: four dedicated to the PEs and one each for the LSU, MXCU, and LCU. To manage diverse workloads, instructions for multiple kernels are hosted in a 10 KiB global IMEM. At kernel onset, these instructions are dispatched from the global IMEM to the private local IMEMs of each specialized element and PE.

\paragraph*{DISCO-CGRA optimizations}

DISCO enhances the baseline VWR2A microarchitecture by targeting the energy-efficiency and data-movement bottlenecks that arise in signal processing and edge-AI workloads. These optimizations affect the PEs and control logic, improving support for high-throughput kernels such as convolutions and matrix operations.

On the processing side, the PE Instruction Set Architecture (ISA) was expanded in DISCO to address the computational demands of edge computing. The ALUs were augmented with a single-cycle MAC operation, providing a performance boost for essential kernels such as GEMM with minimal area overhead. Furthermore, the architecture was enhanced to support SIMD 
operations. This allows a 32-bit word retrieved from a VWR to be processed as two 16-bit operands, effectively doubling the throughput for applications where lower precision is sufficient, such as quantized neural networks.

Finally, the control plane underwent in DISCO a major redesign to reduce power and area overhead. In the original VWR2A, the global IMEM was implemented as a standard-cell memory using sequential logic (flip-flops) to facilitate single-cycle instruction transfers. However, storing 512 instruction words for seven specialized elements required nearly 82,000 flip-flops, leading to excessive static and dynamic power consumption. To mitigate this, the sequential logic was replaced with SRAM macros. Although this necessitated a redesign of the IMEM controller to manage the multi-cycle dispatch at kernel onset, the impact on overall performance is negligible. Because instructions are loaded only once per kernel and typical kernels consist of only a few dozen instructions, this trade-off yields a significant gain in area and power efficiency without compromising execution speed.

% ------------------------------------------------
%             IV. EXPERIMENTAL SETUP
% ------------------------------------------------
\section{Experimental setup}
\label{sec:exp_setup}
% ------------------------------------------------
%             IV.A KERNEL IMPLEMENTATION
% ------------------------------------------------
\subsection{Matrix Multiplication Kernel implementation}
\label{sec:kernel_imp}

This section describes the implementation and mapping strategies for the matrix multiplication kernel on both the OE and DISCO architectures. We consider the operation $C = A \times B$, where the input matrices $A \in \mathbb{R}^{M \times K}$ and $B \in \mathbb{R}^{K \times N}$ produce an output matrix $C \in \mathbb{R}^{M \times N}$. The following subsections detail how the computation is scheduled and how the specific architectural features of each CGRA---such as the VWRs and the SPM in DISCO---are leveraged to optimize data movement and maximize throughput.

\subsubsection{Implementation for OpenEdgeCGRA}
The implementation for OE follows a strictly homogeneous approach, utilizing an output-stationary (OS) dataflow. To match the physical dimensions of the $4 \times 4$ PE mesh, the output matrix is partitioned into corresponding tiles, where the innermost loop executes parallel dot-product computations. A comprehensive breakdown of this implementation is provided in \cite{wang2026exploitingpreoptimizedkernelspolyhedral}.

A critical challenge in this homogeneous design is the limited local memory available within each PE. Unlike heterogeneous architectures with dedicated structures, OE must store kernel configuration parameters—such as base addresses and matrix dimensions—within the same register files used for computation. To mitigate this resource scarcity, configuration values are mapped diagonally across the PEs grid. This specific layout allows each row and column to access metadata for broadcast while minimizing the footprint on the mesh.

However, this reliance on general-purpose registers for control data creates a direct trade-off: as the number of configuration parameters increases, the available register capacity for operand buffering decreases. This register saturation limits the potential for aggressive loop unrolling and data buffering, inherently reducing the kernel's operational efficiency compared to architectures with specialized control hardware. Furthermore, the lack of dedicated control units often forces the allocation of several PEs exclusively for loop management or address generation, effectively withdrawing them from the computational pool and reducing the overall functional parallelism of the mesh.

\subsubsection{Implementation for DISCO-CGRA}

\begin{figure*}[t]
    \centering
    \includegraphics[width=0.8\textwidth]{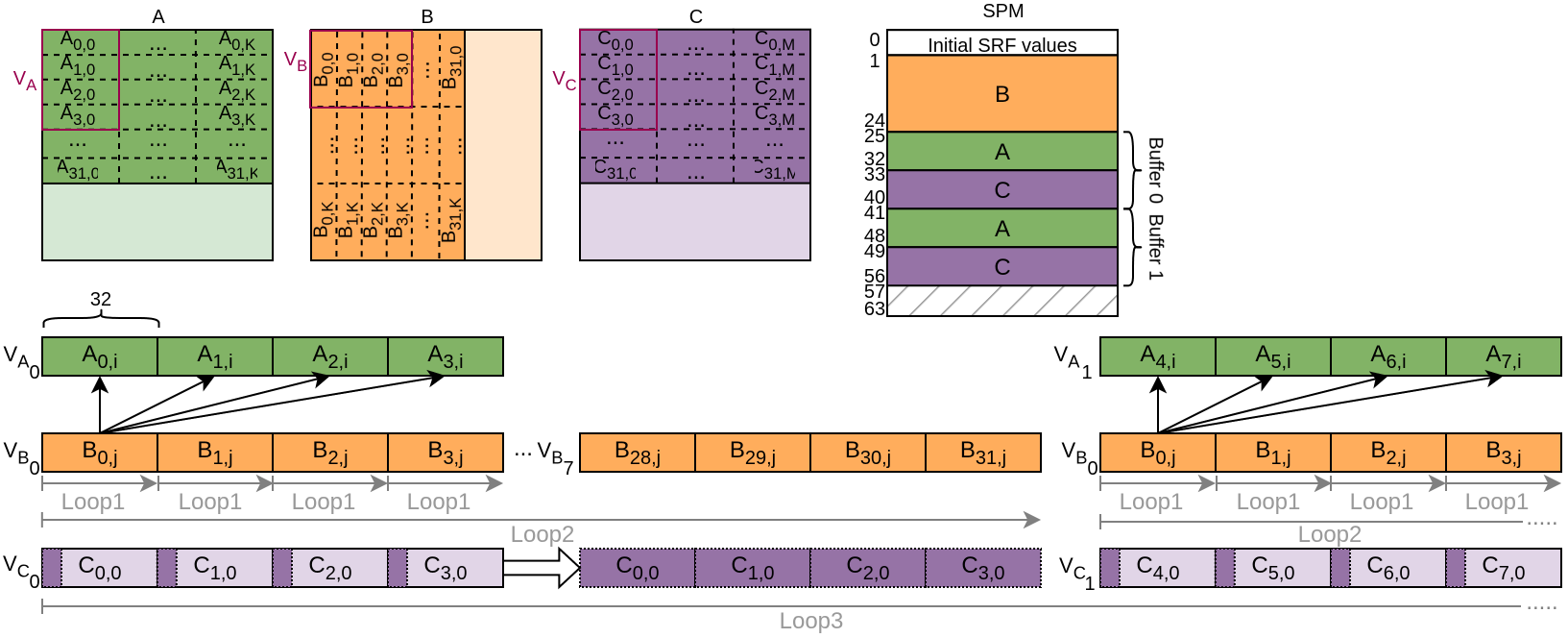}
    \caption{Mapping of a matrix multiplication kernel ($C = A \times B$) onto the DISCO architecture. The diagram illustrates the SPM organization and VWR placement, where vectors $A_{i,j}, B_{i,j}$, and $C_{i,j}$ represent 32-element segments. Darker colors indicate the active data blocks for the current iteration, while dotted rectangles represent partially computed results. The layout highlights the row-wise storage for $A$ and $C$ versus the column-wise (transposed) storage for $B$.}
    \label{fig:disco_mmul}
\end{figure*}

DISCO's GEMM kernel is optimized for blocks of size $32 \times 32$, to fill the 128-element capacity of the VWRs, where each of the four PEs manages a 32-element segment.

\paragraph{Data Organization and Memory Hierarchy}
The kernel is fully parameterized to support flexibility in matrix dimensions and memory offsets. At execution onset, loop boundaries and initial addresses are fetched from the SPM and loaded into the SRF, making them globally accessible to all specialized units. This parameterization enables a double-buffering strategy, allowing the system to overlap data transfers between main memory and the SPM with active computation, effectively hiding memory latency. To maximize throughput, the data layout in the VWRs is specialized by matrix type. Matrices A and C are stored row-wise, where a single VWR holds 32 elements from four different rows. Conversely, matrix B is stored column-wise. To accommodate the interconnect and vector access patterns, B is transposed before being loaded into the SPM. Because of the complexity of this transposition, the kernel is designed to maximize the reuse of B once it is resident in the SPM.

\paragraph{Kernel Execution}
As illustrated in Figure~\ref{fig:disco_mmul}, the kernel execution is orchestrated through three nested loops managed by the LCU. The hierarchy is structured such that the outer loop (L3) controls the slice of C currently being computed, which dictates the corresponding rows of A to be loaded into the VWRs. The intermediate loop (L2) manages the columns of B being processed, triggering the LSU to update the current B values in the VWR with the next set of data stored in the SPM. Finally, the inner loop (L1) iterates through the K dimension; in each cycle, an element of B is broadcast to the PEs, which perform MAC operations against the rows of A. During this process, the LSU handles all data movement between the SPM and VWRs, while the MXCU manages the specific index pointers for VWR segments and SRF lookups. The PEs compute partial sums in their local registers, and only once a full dot product is completed the result is accumulated on the output VWR.

\paragraph{Scalability and Load Balancing}
For matrices exceeding $32 \times 32$, tiling is performed by the host CPU. The output matrix C is decomposed into kernel-sized blocks, and the double-buffering mechanism ensures the next tile is ready as soon as the current block finishes. To ensure maximum efficiency, the workload is distributed across both independent CGRA columns. The system is balanced such that both columns execute symmetric workloads, ensuring they reach the completion synchronization point simultaneously.

\subsubsection{Comparative Analysis of Implementation Overhead}

\begin{figure*}[t]
    \centering
    \includegraphics[width=0.9\textwidth]{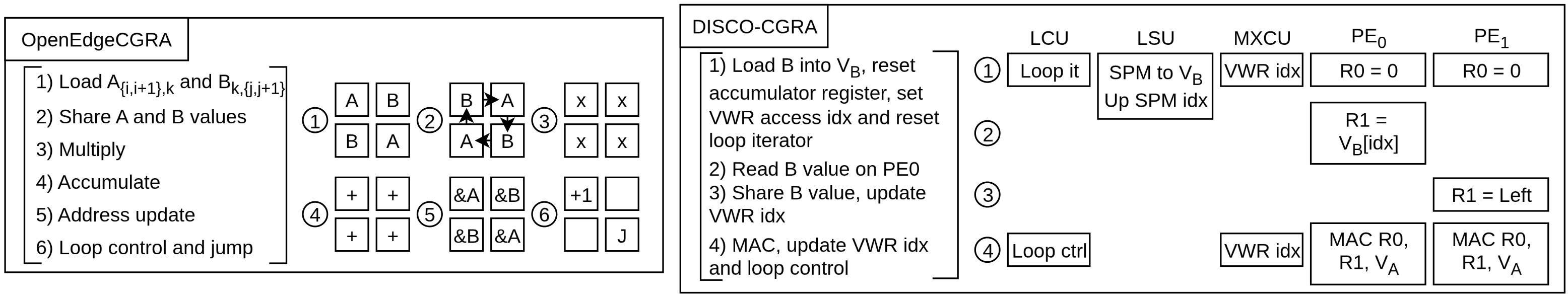}
    \caption{Comparative implementation of the innermost loop for a matrix multiplication kernel using OE and DISCO mapping strategies on a simplified $2 \times 2$ mesh CGRA.}
    \label{fig:DISCO_OE_inner_loop}
\end{figure*}

This section provides a preliminary comparative analysis of the implementation efficiency by evaluating the instruction count and the theoretical cycle overhead within the innermost loop. This analytical study quantifies the impact of DISCO's hardware heterogeneity against the homogeneous structure of OE before proceeding to the detailed experimental characterization. A full evaluation of execution time, power consumption, and energy efficiency is discussed in depth in Section~\ref{sec:results}.

The architectural differences between the two approaches are illustrated in Figure \ref{fig:DISCO_OE_inner_loop}, which depicts the innermost loop structure for a matrix multiplication kernel mapped onto a simplified $2 \times 2$ mesh CGRA. The impact of architectural heterogeneity and the inclusion of an SPM are evident in the resulting instruction overhead. Regarding the contrast between homogeneity and heterogeneity, the OE approach forces all processing elements to simultaneously manage both computation and control flow due to its homogeneous nature. Conversely, DISCO leverages heterogeneity by assigning specialized tasks to specific PEs, allowing control tasks to be executed in parallel with computation. This architectural choice reduces the innermost loop complexity from 6 instructions per PE in OE to only 4 instructions per PE in DISCO. Across the entire fabric, this translates to a massive reduction in parallel instruction overhead: while the homogeneous OE architecture requires a total of 96 instructions across its 16 PEs (4 columns of 4 PEs), DISCO's heterogeneous fabric requires a total of only 56 instructions across its 14 units (2 columns, each executing 4 PE and 3 specialized unit instructions simultaneously). Furthermore, the presence of an SPM in DISCO significantly streamlines data handling. In the OE implementation, generating memory addresses for storing partial sums and preparing subsequent iterations requires 8 instructions in the middle loop. In contrast, DISCO eliminates complex address generation arithmetic entirely. The LSU directly handles the scratchpad line indices while the MXCU manages the sub-line element indexing, reducing the process to only 2 instructions.

% ------------------------------------------------
%                IV.B END-TO-END APP
% ------------------------------------------------
\subsection{End-to-End Application: Seizure Detection Transformer}
\label{sec:transformer}

To evaluate the performance of both CGRA architectures in a real-world biomedical scenario, we employ a modified 4-layer ViT \cite{dosovitskiy2020image} specifically optimized for the TSD task from 20-channel EEG signals \cite{ma2023tsd, amirshahi2024fetch}.

\begin{figure*}[t]
    \centering
    
    % --- Fila 1: Esquema General (Ancho total) ---
    \subfloat[Overview of the end-to-end seizure detection transformer architecture.]{
        \includegraphics[width=0.98\textwidth]{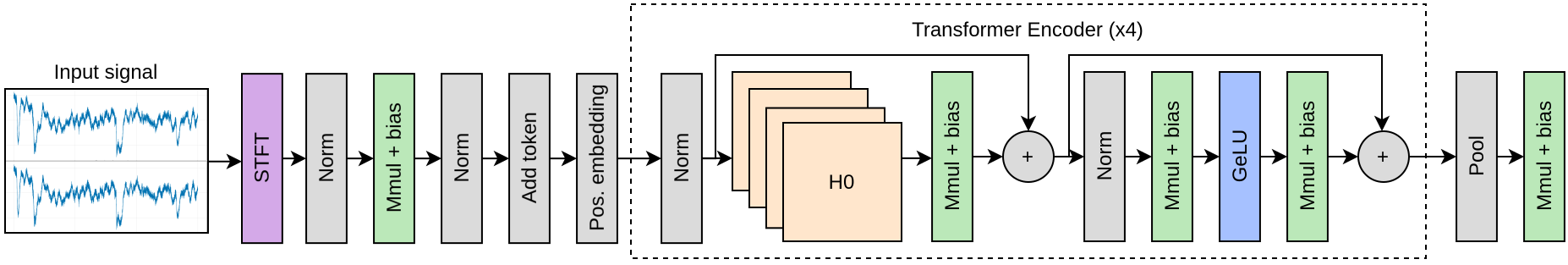}
        \label{fig:transf_scheme_general}
    }

    \vspace{4mm}

    % --- Fila 2, Izquierda: Head Scheme (Ocupa la mitad del ancho) ---
    \begin{minipage}{0.48\textwidth}
        \centering
        \subfloat[Detailed operations of a MHSA block.]{
            \includegraphics[width=0.85\textwidth]{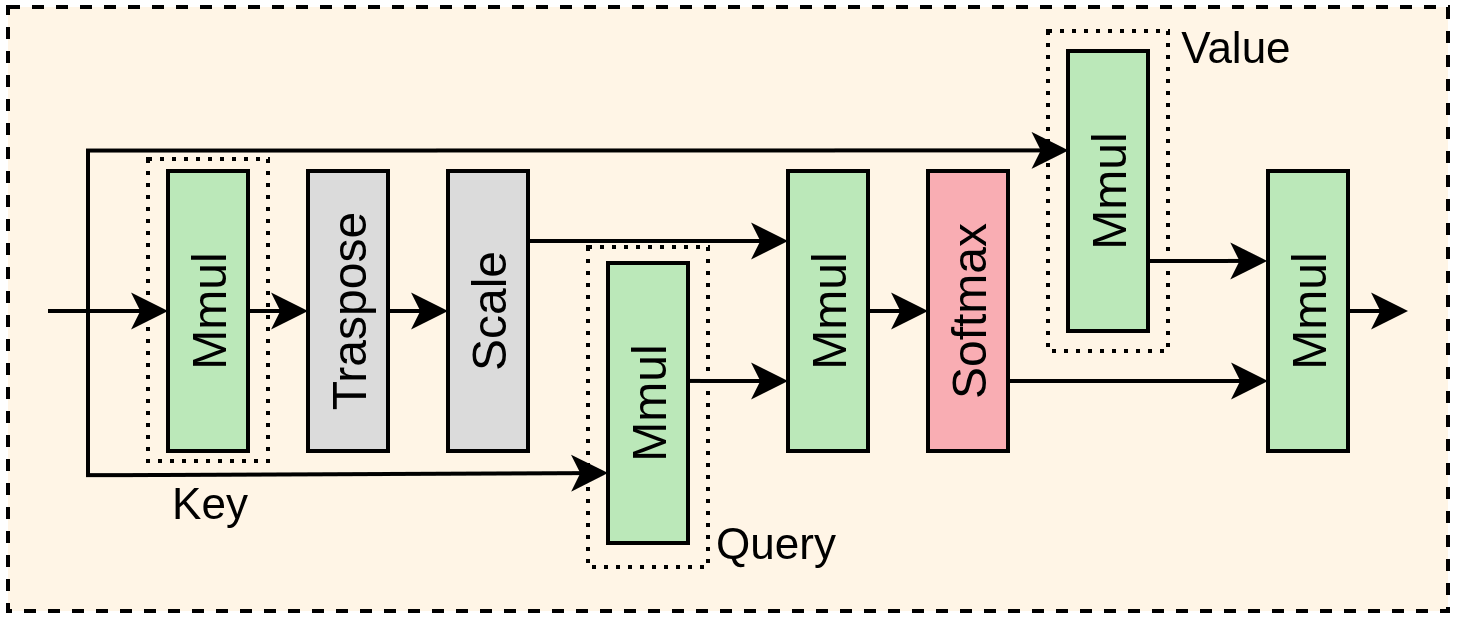}
            \label{fig:transf_head_scheme}
        }
    \end{minipage}
    \hfill
    % --- Fila 2, Derecha: Los dos Pie Charts horizontales (Ocupa la otra mitad) ---
    \begin{minipage}{0.48\textwidth}
        \centering
        \subfloat[Transformer app profiling with ConSmax and ConSmax after row-wise common factor (cf) optimization.]{
            \includegraphics[width=0.9\textwidth]{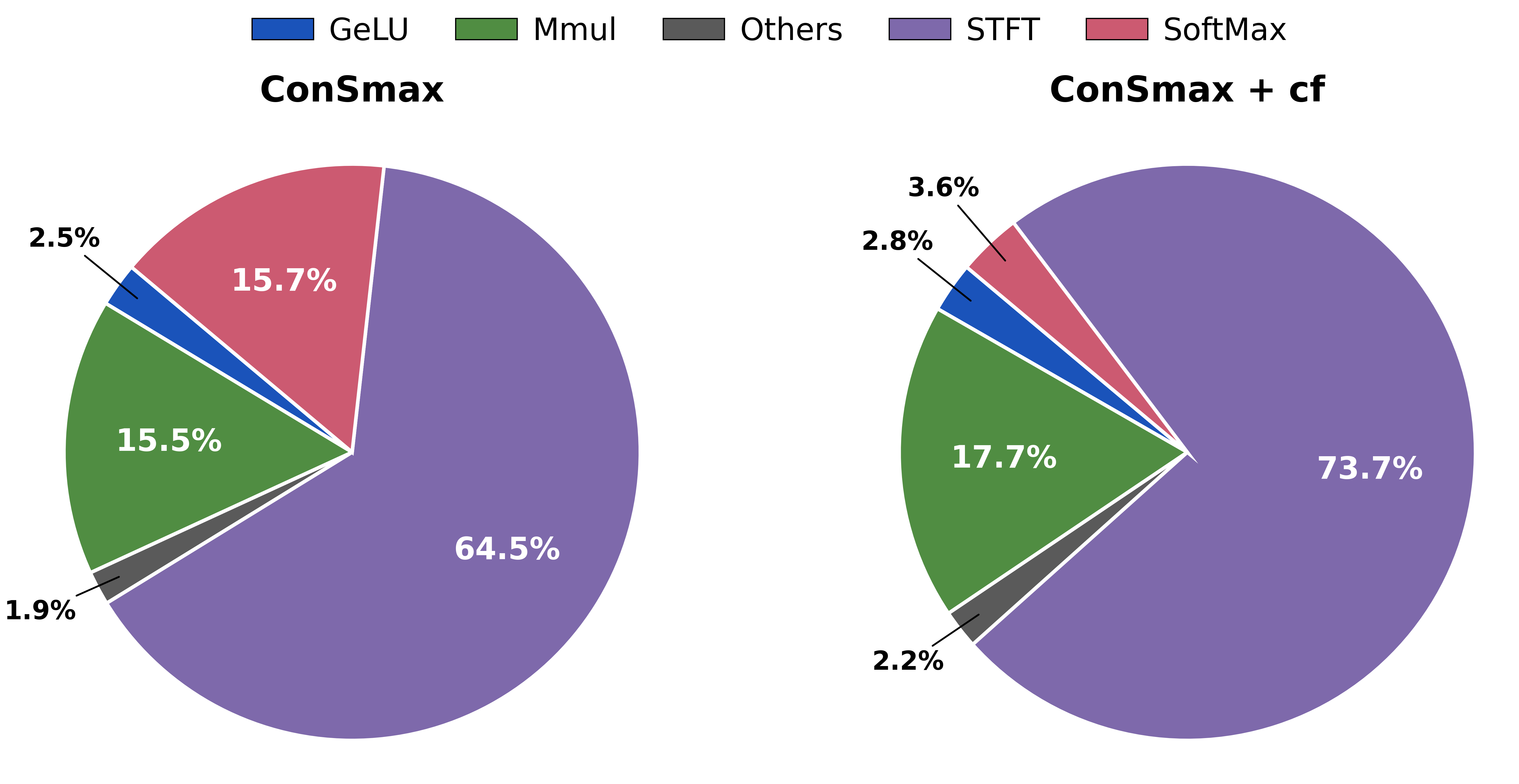}
            \label{fig:piechart_transformer}
        }
    \end{minipage}

    \vspace{1mm}

    \caption{Detailed architectural overview and execution profiling of the proposed seizure detection transformer: (a) full end-to-end pipeline highlights computing-intensive and non-linear operations; (b) internal operations of the MHSA block; (c) impact of ConSmax with row-wise common factor (cf) optimizations on the computational workload distribution. Color coding of workload kernels based on computational intensity and execution target. Computational kernels accelerated on the CGRA are colored green (Matrix Multiplication) and purple (STFT). Non-linear kernels executed on the CPU are highlighted in blue (GeLU) and pink (Softmax) due to their significant share of total execution time. Remaining host-side tasks are colored gray, indicating standard CPU execution.}

    \label{fig:trasnformer_overview_profiling}
\end{figure*}

\subsubsection{Model Architecture and Signal Processing}
The application pipeline, illustrated in Figure~\ref{fig:transf_scheme_general}, begins with the transformation of 12-second EEG segments using the Short-Time Fourier Transform (STFT). These time-frequency features are flattened into patches and mapped through an embedding layer. 

The core of the model consists of four Transformer Encoder layers. Each layer integrates a Multi-Head Self-Attention (MHSA) mechanism, where the input is decomposed into Query ($Q$), Key ($K$), and Value ($V$) matrices, shown in Figure~\ref{fig:transf_head_scheme}. The Softmax function is applied within this MHSA block to compute the attention scores. The output of the attention is then combined with the residual connection (skip connection) before proceeding to the final classification head to produce the clinical prediction.

\subsubsection{Profiling and Kernel Selection}
Detailed profiling of the transformer application, as illustrated in Figure~\ref{fig:piechart_transformer}, reveals that three operations dominate the computational overhead: the STFT accounting for 64.5\% of the total execution cycles, followed by the softmax activation with 15.7\%, and Matrix Multiplication (Mmul) with 15.5\%. 

While the FFT kernels within the STFT and Mmul are highly data-parallel and suitable for CGRA offloading, the softmax presents a challenge. Its reliance on transcendental functions, floating-point arithmetic, and high-precision divisions makes it unsuitable for the integer-based ALUs of the targeted CGRAs. Consequently, the softmax is executed on the host processor's CPU.

\subsubsection{Softmax Optimization and Mathematical Formulation}
To reduce the overhead of softmax computation in attention scoring, we employ a hardware-friendly constant softmax approximation based on a three-coefficient Taylor expansion, following the approach proposed in~\cite{liu2024consmax, taji2025medea}. The proposed approach formulates a softmax constant approximation as follows. For an individual element $S_{i,j}$ of the score matrix $S \in \mathbb{R}^{N \times K}$, the approximation is defined as:

\begin{equation*}
    \text{ConSmax}(S_{i,j}) = \frac{e^{S_{i,j}} - \beta}{\gamma} = C_i \times e^{S_{i,j}}
\end{equation*}

where $C_i = -e^{\beta}/\gamma$. In a baseline implementation, this operation is applied individually to every element of the matrix. However, as shown in the following equation, the constant $C_i$ is identical for all elements within the same row:

\begin{equation*}
\text{ConSmax}(S) = 
\begin{bmatrix} 
C_0 e^{S_{0,0}} & \dots & C_0 e^{S_{0,K-1}} \\ 
C_1 e^{S_{1,0}} & \dots & C_1 e^{S_{1,K-1}} \\ 
\vdots & \ddots & \vdots \\ 
C_{N-1} e^{S_{N-1,0}} & \dots & C_{N-1} e^{S_{N-1,K-1}} 
\end{bmatrix}.
\end{equation*}

By extracting $C_i$ as a common factor for each row, we significantly optimize the computation. Instead of performing the division and parameter calculations for every single entry, we compute the vector of constants once per row and then multiply it by the exponentiated matrix:

\begin{equation*}
\text{ConSmax}(S) = 
\begin{bmatrix} 
C_0 \\ 
C_1 \\ 
\vdots \\ 
C_{N-1} 
\end{bmatrix} 
\times 
\begin{bmatrix} 
e^{S_{0,0}} & \dots & e^{S_{0,K-1}} \\ 
e^{S_{1,0}} & \dots & e^{S_{1,K-1}} \\ 
\vdots & \ddots & \vdots \\ 
e^{S_{N-1,0}} & \dots & e^{S_{N-1,K-1}} 
\end{bmatrix}.
\end{equation*}

As shown in Figure ~\ref{fig:piechart_transformer}, this optimization reduces the Softmax overhead by a factor of 5, bringing its impact down to only 3.6\% of the total execution cycles. This rebalances the system's profile, leaving two primary dominant kernels for acceleration: STFT with 73.7\% and Mmul with 17.7\%, which together account for 91.4\% of the total application time.

% ------------------------------------------------
%                     IV.C SETUP
% ------------------------------------------------
\subsection{Measurement setup}
\label{sec:setup}

To evaluate the key performance metrics of the OE, DISCO, and baseline CV32E40P architectures, logic synthesis was performed using Cadence Genus \cite{noauthor_genus_nodate} targeting the TSMC 16nm FinFET (SVT) technology node. Timing analysis was conducted under the worst-case PVT corner (0.72 V, 40°C), while power analysis utilized the typical corner (0.8 V, 25$^\circ$C). Through iterative synthesis, the maximum achievable clock frequency for each design was determined. DISCO achieved a maximum frequency of 200 MHz, while the OE architecture reached 700 MHz.

Consequently, the architectures were evaluated under two scenarios: first, a fair power and area comparison was conducted at an iso-frequency of 200 MHz, matching the maximum capability of the slower design; second, each architecture was evaluated at its respective maximum frequency (200 MHz for DISCO and 700 MHz for OE) to determine application execution time under peak operational speeds. Following synthesis, cycle-accurate post-synthesis simulations were executed for each architecture across the benchmark kernels to capture precise workload switching activity. Finally, Synopsys PrimePower \cite{noauthor_primepower_nodate} utilized these switching activities alongside the synthesized frequencies and execution cycles to estimate the final execution time and energy consumption for each specific workload.

% ------------------------------------------------
%                V. RESULTS
% ------------------------------------------------
\section{Results}
\label{sec:results}

In this section, we analyze the performance, power, and energy characteristics of the DISCO and OE architectures. The evaluation is divided into three parts to cover different aspects of architectural efficiency. First, the Matrix Multiplication Analysis in Section~\ref{sec:mmul_analysis} focuses on data-intensive workloads using benchmarks from the PolyBench~\cite{pouchet2012polybench} suite, specifically \textit{mmul}, \textit{gemm}, \textit{2mm}, and \textit{3mm}. This analysis evaluates how data reuse patterns in multi-stage operations leverage the internal SPM to optimize throughput. Second, in Section~\ref{sec:end_to_end}, we evaluate the full integration of the seizure detection transformer model described in Section~\ref{sec:transformer} to explore the trade-offs of executing a complete biomedical pipeline. Finally, in Section~\ref{sec:sota_comp}, we assess the performance of the FFT across several SoA CGRAs to demonstrate that both DISCO and OE achieve performance levels consistent with or exceeding current academic and industrial benchmarks.

% ------------------------------------------------
%                V.A MMUL ANALYSIS
% ------------------------------------------------
\subsection{Matrix Multiplication analysis}
\label{sec:mmul_analysis}

\begin{table}[ht]
\centering
\caption{Memory traffic comparison (KiB) between OE and DISCO for matrix multiplication benchmarks of different sizes.}
\label{tab:memory_traffic}
\begin{tabular}{|l|rr|rr|}
\hline
 & \multicolumn{2}{c|}{\textbf{OE}} & \multicolumn{2}{c|}{\textbf{DISCO}} \\ \cline{2-5} 
\textbf{Benchmark} & \multicolumn{1}{c}{\textbf{Loaded}} & \multicolumn{1}{c|}{\textbf{Stored}} & \multicolumn{1}{c}{\textbf{Loaded}} & \multicolumn{1}{c|}{\textbf{Stored}} \\ \hline
mmul 16  & 8    & 1  & 8   & 4 \\
mmul 32  & 64   & 4  & 8   & 4 \\
mmul 64  & 512  & 16 & 64  & 16 \\
mmul 128 & 4096 & 64 & 512 & 64 \\ \hline
\end{tabular}
\end{table}

\begin{figure*}[t]
    \centering
    
    \begin{minipage}{0.75\textwidth}
        \centering
        \subfloat[Total execution cycles (bars), absolute execution time (lines with markers), and time speedup (annotated above the time markers).]{
            \includegraphics[width=0.9\textwidth]{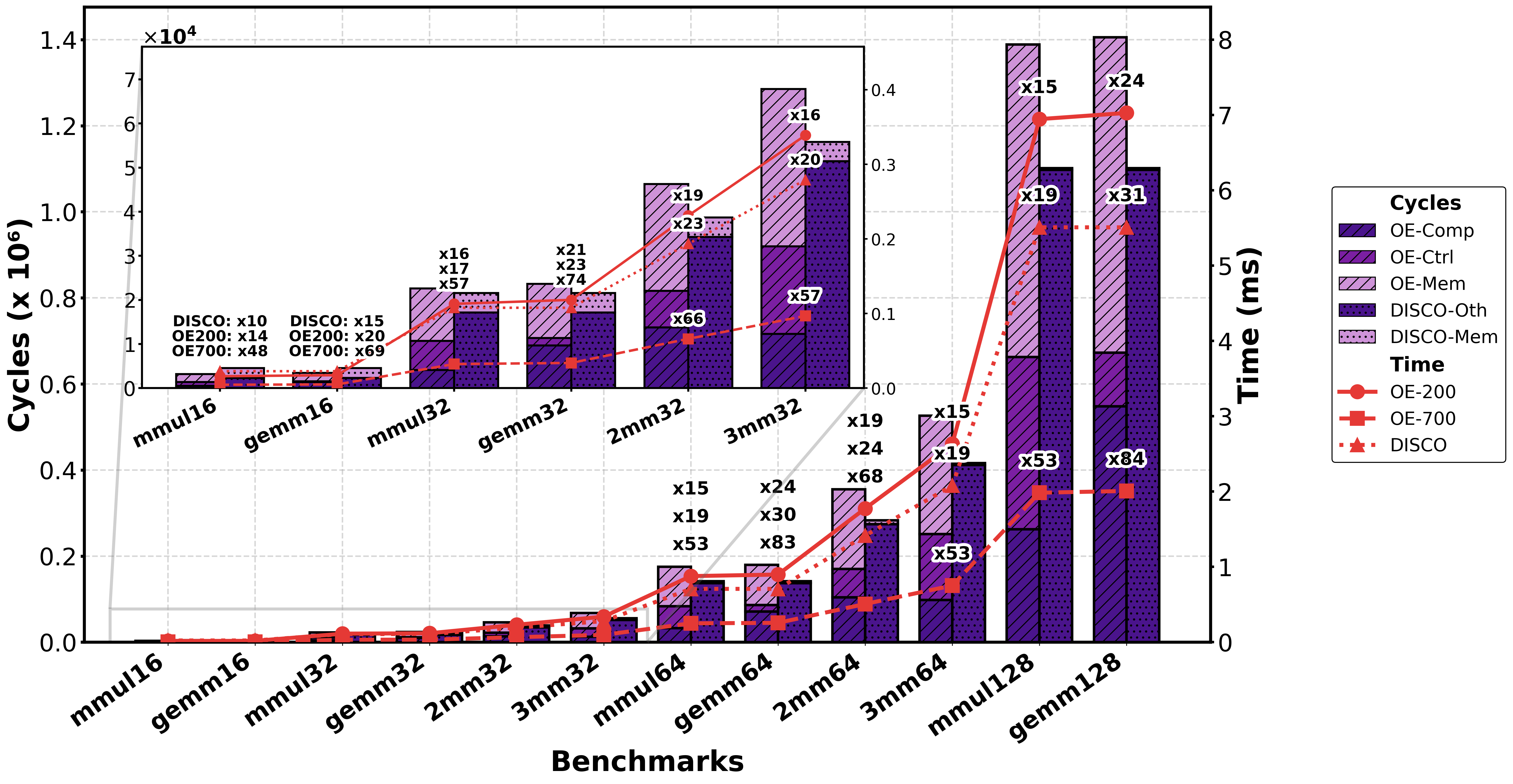}
            \label{fig:time_mat_bench}
        } 
    \end{minipage}
    
    \begin{minipage}{0.49\textwidth}
        \centering
        \subfloat[Energy consumption (bars) and ratio w.r.t CPU (markers)]{
            \includegraphics[width=\textwidth]{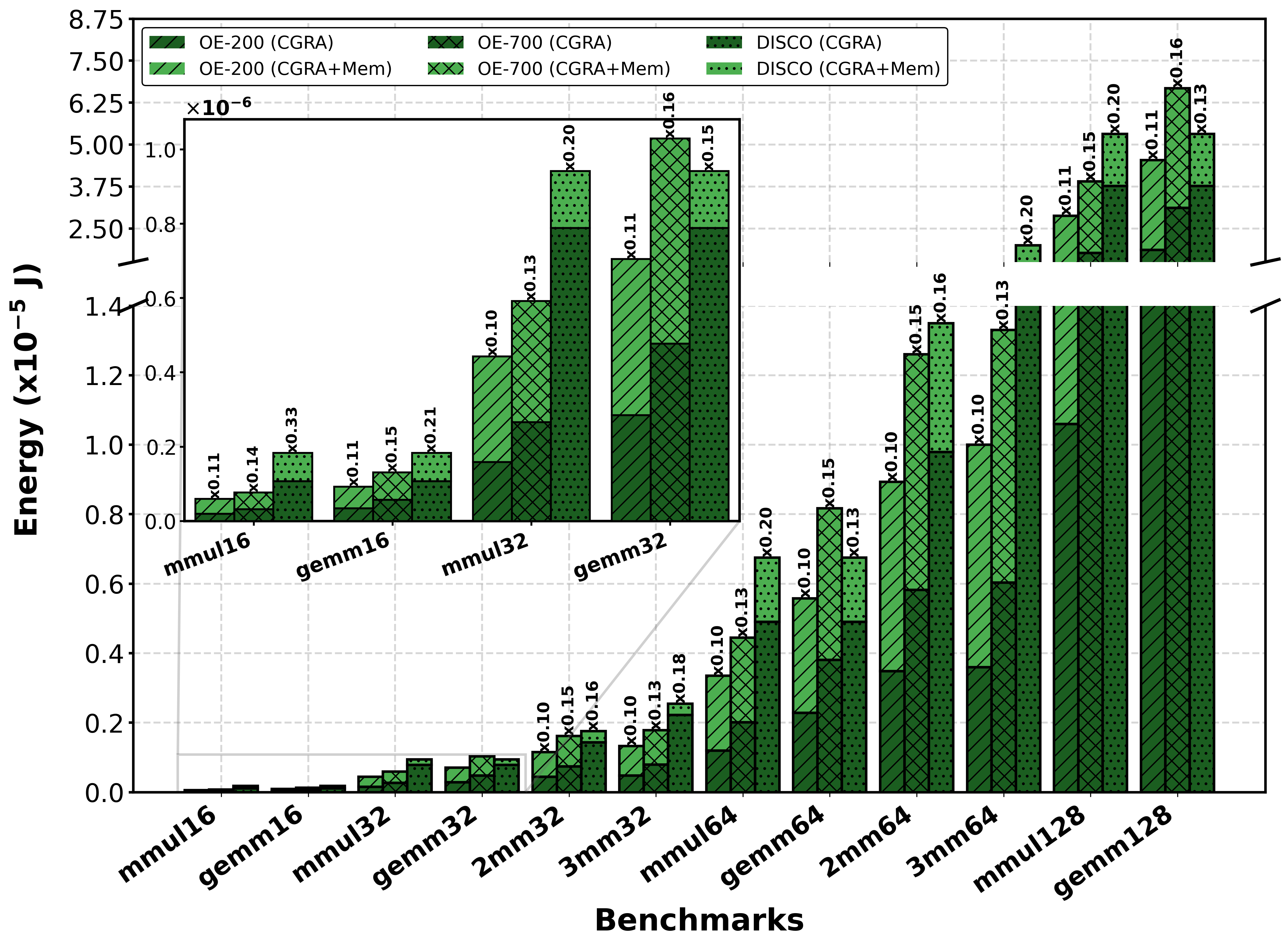}
            \label{fig:energy_mat_bench}
        }
    \end{minipage}
    \hfill
    \begin{minipage}{0.49\textwidth}
        \centering
        \subfloat[Average power dissipation (bars) and relative power ratio w.r.t. CPU (markers).]{
            \includegraphics[width=\textwidth]{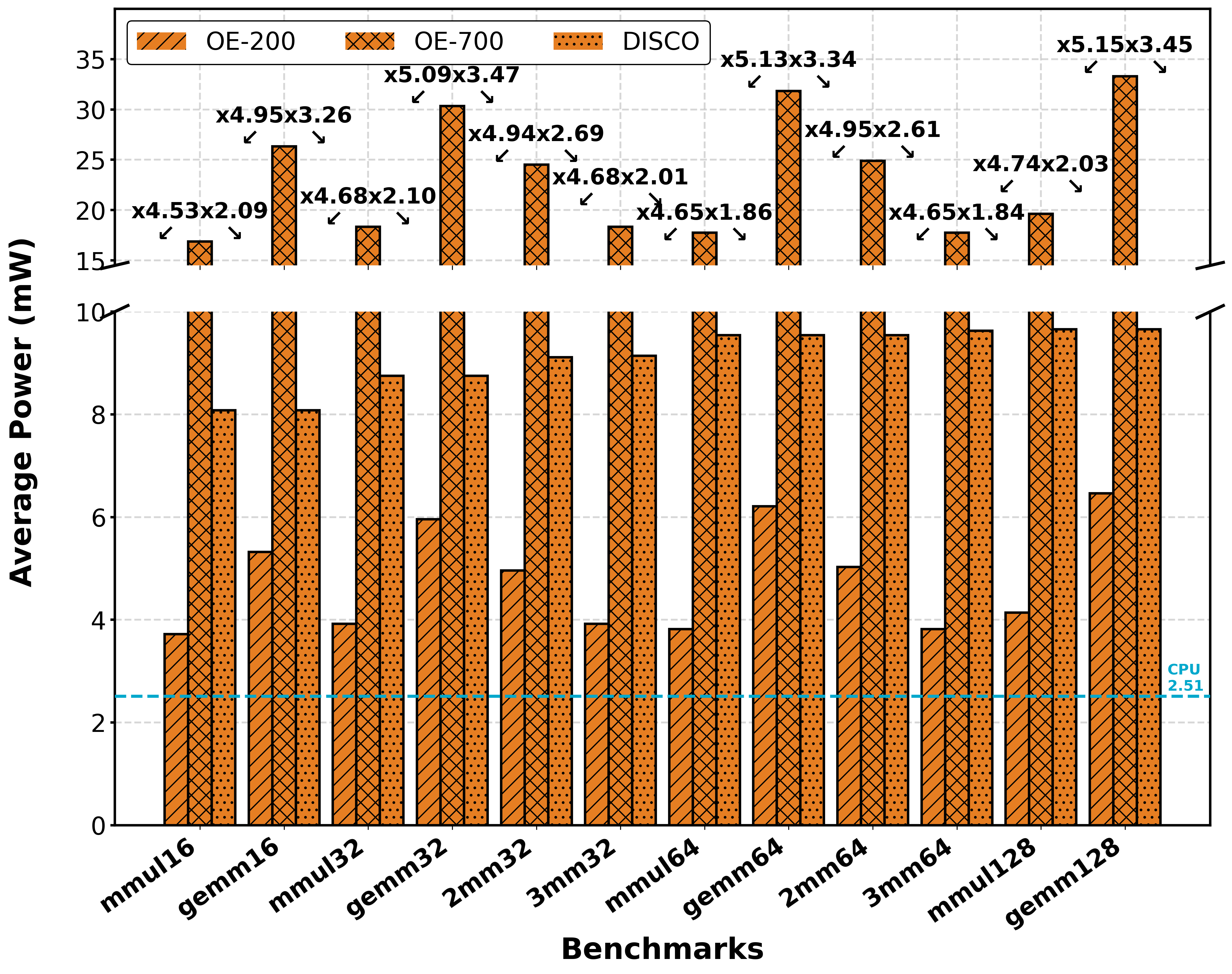}
            \label{fig:power_mat_bench}
        }
    \end{minipage}
    \caption{Experimental evaluation of DISCO and OE architectures across a set of Polybench~\cite{pouchet2012polybench} matrix multiplication kernels. (a) Execution performance in cycles and time, including a comparison against a baseline CPU. (b) Total energy consumption, where ``CGRA'' represents the energy of the accelerator core alone, and ``CGRA+Mem'' includes the energy consumed by the memory subsystem to service loads/stores. (c) Average power dissipation. For all bar charts, lower values indicate better efficiency or performance. For graphs utilizing markers, larger values indicate better performance for execution time speedups, whereas lower values indicate superior efficiency for energy consumption ratios and power dissipation profiles.}
    \label{fig:benchmarks_results}
\end{figure*}

We compare OE and DISCO through a set of Polybench~\cite{pouchet2012polybench} benchmarks to assess how specialized PEs and SPM influence overall system throughput, average power and energy consumption.

Specifically, the impact of the SPM in DISCO is directly visible in the volume of data transferred between main memory and the accelerator. Since both OE and DISCO use the same OS approach for $mmul$, they store outputs equally often except for the small data sizes where DISCO employs padding, but the OS strategy requires repeated access to input data during execution. In the case of DISCO, the data that needs to be accessed frequently is kept within the SPM, where it can be easily reused. In contrast, OE lacks this local storage, forcing the system to transfer the same data repeatedly from main memory. The impact of this architectural difference on the $mmul$ kernel (representative of all benchmarks) is quantified in Table \ref{tab:memory_traffic}, which shows that memory loads in OE are significantly higher, with total traffic exceeding that of DISCO by a factor of 8. Ultimately, these redundant memory accesses translate into higher execution stall cycles and increased I/O activity, which significantly degrades throughput and inflates energy consumption.

The performance, energy and power results for the set of Polybench suite—comprising $mmul$, $gemm$, $2mm$, and $3mm$ kernels—are illustrated in Figure \ref{fig:benchmarks_results}. The performance analysis, as illustrated in the breakdown of execution cycles in Figure~\ref{fig:time_mat_bench}, reveals the fundamental efficiency of the DISCO architecture compared to OE. In DISCO, the control and computation cycles are consolidated into the \textit{DISCO-Oth} metric; because this category includes loop control and internal data movements between SPM and VWRs alongside raw execution, its total is larger than OE's standalone control and computation bars. However, DISCO demonstrates near-zero exclusive memory cycles because its double-buffering mechanism effectively masks data transfers. This overhead only becomes visible in small matrix dimensions ($\leq16$) where the SPM cannot be fully utilized to hide the initial latency. Conversely, OE spends a significant portion of its execution cycles on memory transfers and control logic, with its PEs engaged in active computation for only approximately one-third of the total cycles. Consequently, for all benchmarks beyond the dimension $32$ threshold, DISCO achieves lower cycle counts than OE. In terms of absolute execution time, DISCO's performance varies based on the workload characteristics. For the $mmul16$ and $gemm16$ benchmarks, DISCO underperforms due to padding overheads. Specifically, OE-200 and OE-700 achieve speedups of up to $1.5\times$ and $5.25\times$ over DISCO, respectively. Conversely, across all other benchmarks, DISCO is highly competitive with OE-200, achieving a modest speedup ranging from 1.04$\times$ to 1.27$\times$. However, the higher operating frequency of OE-700 allows it to maintain a $2.74\times$ to $3.36\times$ speedup over DISCO on these same workloads.

The impact of these memory access patterns is further evidenced in the energy profile (Figure~\ref{fig:energy_mat_bench}). Notably, DISCO exhibits a significantly reduced ratio of memory-to-cgra energy consumption compared to OE. This efficiency is a direct consequence of the aforementioned SPM integration, which effectively minimizes the frequency of main memory activations required to load values. In terms of absolute energy consumption across the benchmarks, the OE-200 configuration consistently achieves the lowest energy footprint. Conversely, the energy trade-off between OE-700 and DISCO is highly workload-dependent. For the $gemm$ kernels, where memory activations escalate sharply on the OE architecture, OE-700 incurs a higher energy penalty than DISCO. On the other hand, for the $mmul$ kernels, DISCO exhibits higher energy consumption relative to the OE-700 configuration.

Similarly, the average power dissipation profiles illustrated in Figure~\ref{fig:power_mat_bench} reflect these architectural trade-offs. The OE-700 configuration consistently exhibits the highest power dissipation, requiring between $1.84\times$ and $3.47\times$ more power than DISCO across all evaluated benchmarks. Meanwhile, DISCO operates at an intermediate power bracket, exhibiting between $1.46\times$ and $2.52\times$ higher power dissipation than the OE-200. In absolute terms, the OE-200 configuration strictly remains below a $6.5\text{ mW}$ power threshold. In contrast, DISCO's power dissipation ranges between $8\text{ mW}$ and $10\text{ mW}$, while the high-frequency OE-700 configuration peaks significantly higher, fluctuating between $17\text{ mW}$ and $33\text{ mW}$.

% ------------------------------------------------
%                V.B END-TO-END APP
% ------------------------------------------------
\subsection{End-to-end app}
\label{sec:end_to_end}

\begin{figure}[t!]
    \centering
    \begin{minipage}{0.95\columnwidth}
        \centering
        \subfloat[Total execution time (bars) and speedup w.r.t. CPU (markers).]{
            \includegraphics[width=\textwidth]{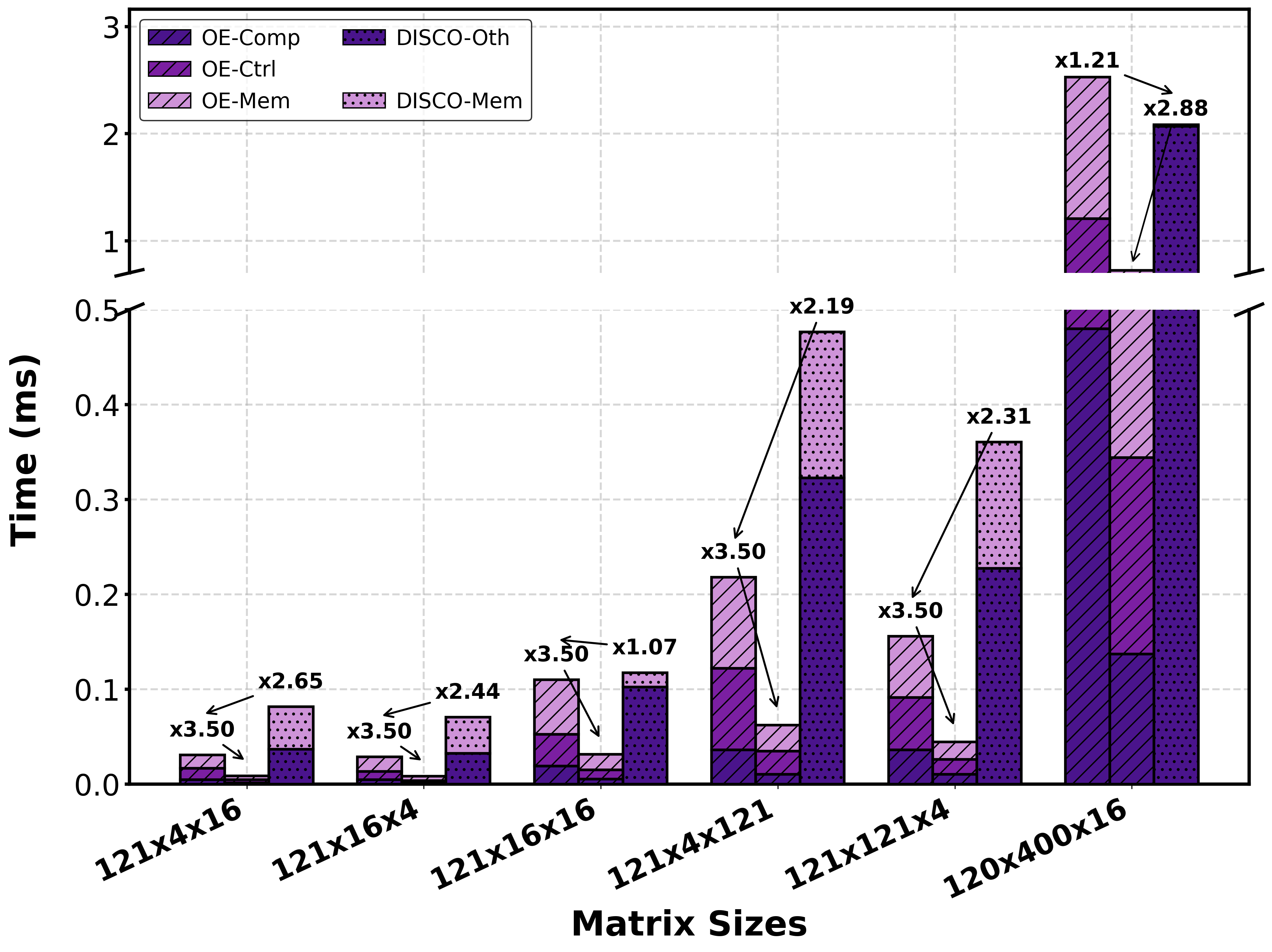}
            \label{fig:time_mat_transf}
        }
    \end{minipage}
    \vspace{0.3cm} % Adds spacing between the stacked subfigures
    
    \begin{minipage}{0.95\columnwidth}
        \centering
        \subfloat[Energy consumption (bars) and ratio w.r.t. CPU (markers).]{
            \includegraphics[width=\textwidth]{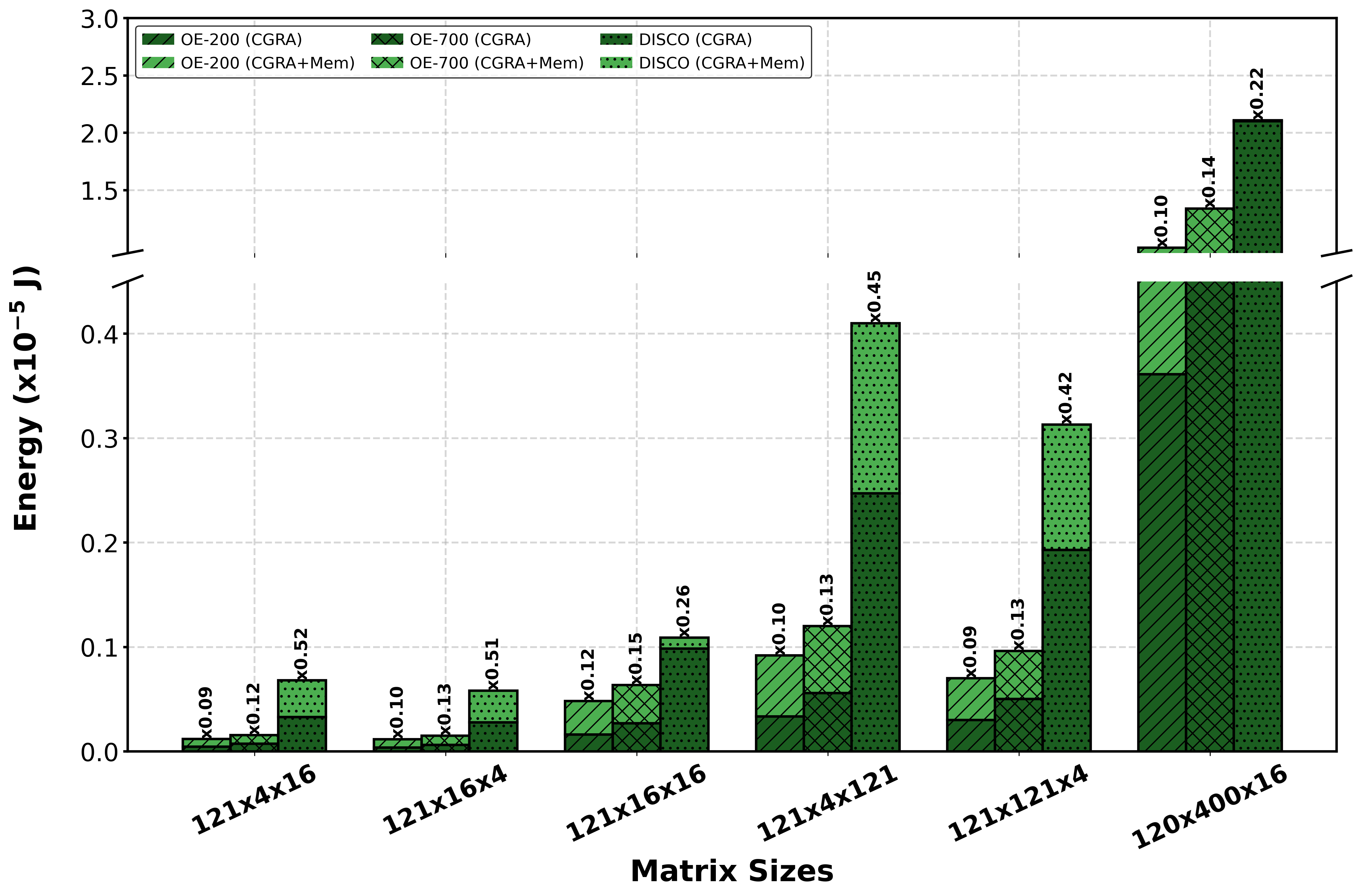}
            \label{fig:energy_mat_transf}
        }
    \end{minipage}
    \vspace{0.3cm}
    
    \begin{minipage}{0.95\columnwidth}
        \centering
        \subfloat[Average power dissipation (bars) and ratio w.r.t. CPU (markers).]{
            \includegraphics[width=\textwidth]{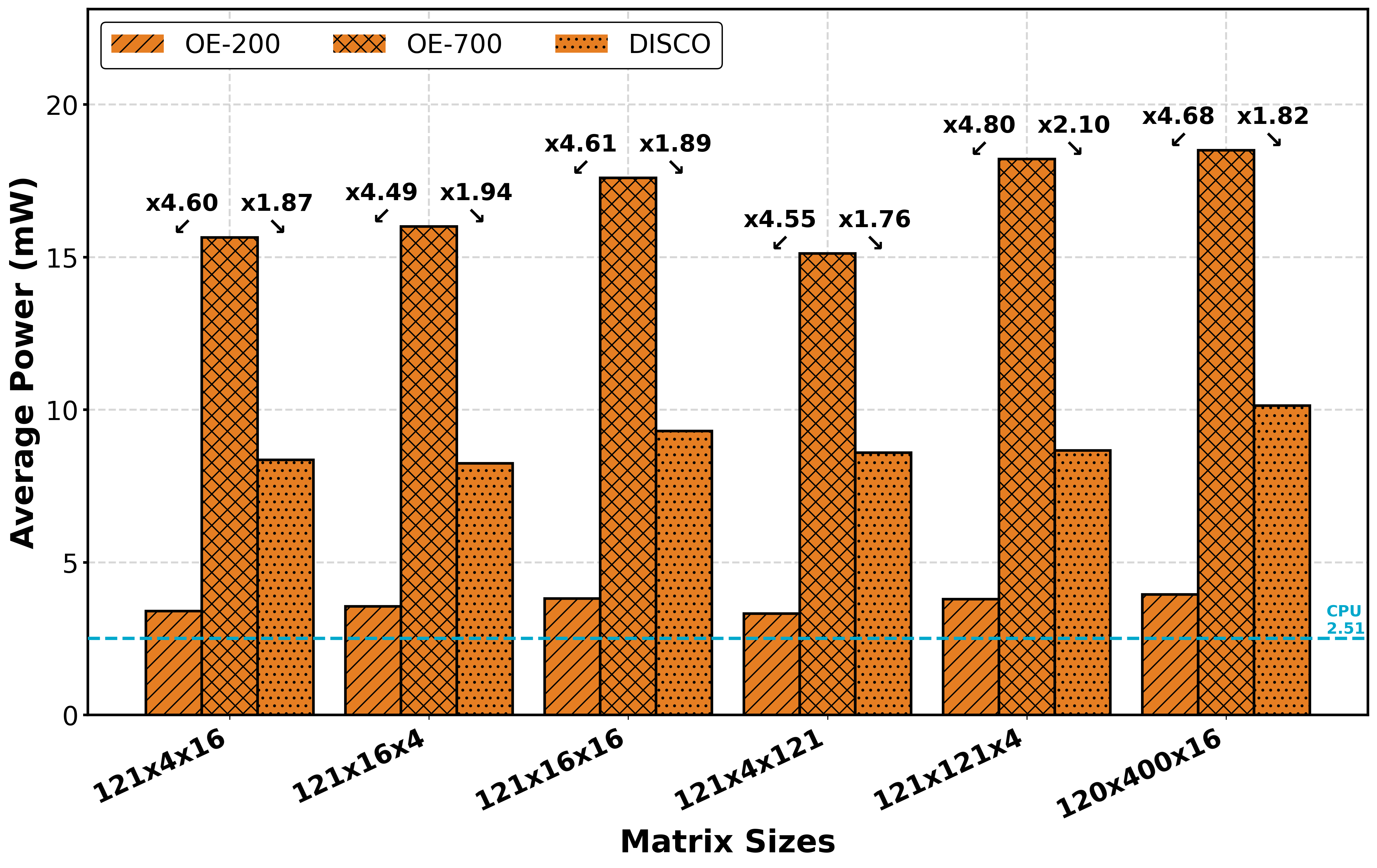}
            \label{fig:power_mat_transf}
        }
    \end{minipage}

    \caption{Experimental evaluation of the mmul kernel across different matrix sizes present in the TSD pipeline: (a) execution time profiling with speedup annotations; (b) energy consumption and its relative ratio compared to the CPU baseline; (c) average power dissipation and its relative ratio compared to the CPU baseline. For all bar charts, lower values indicate better efficiency or performance. For graphs utilizing markers, larger values indicate better performance for time-related metrics, whereas lower values indicate superior efficiency for energy consumption and power dissipation.}
    \label{fig:matrix_sizes_transformer_eval}
\end{figure}

To further analyze the performance dynamics of the TSD pipeline, we evaluate the individual matrix multiplications within the transformer layers, as illustrated in Figure~\ref{fig:matrix_sizes_transformer_eval}. In terms of execution time, OE-700 consistently achieves the lowest latency across all configurations. OE-200 also outperforms DISCO in nearly all scenarios, with the sole exception of the largest matrix operation, where DISCO exhibits a marginal advantage and OE-200 requires $1.21\times$ more execution time. Energy consumption trends closely mirror these latency behaviors, with DISCO demanding the highest energy footprint. Notably, as the matrix dimensions scale, DISCO increasingly mitigates memory-related energy overheads due to the continuous activation of the CGRA. Overall, OE-200 remains the most energy-efficient configuration, closely followed by OE-700. In comparison, DISCO consumes substantially more energy, requiring $2.11\times$ to $5.72\times$ more energy than OE-200, and $1.57\times$ to $4.37\times$ more than OE-700. Despite this overhead, all CGRA configurations maintain a massive efficiency advantage over the baseline CPU. For instance, DISCO's energy consumption is merely $0.13\times$ to $0.52\times$ that of the CPU. Regarding power dissipation, the configurations follow the strict ordering of $\text{OE-200} < \text{DISCO} < \text{OE-700}$, a trend that aligns consistently with our observations across the broader benchmark suite.

\begin{figure*}[t!]
    \centering
    % ---  STFT ---
    \subfloat[Execution cycles (bars) and speedup relative to a baseline CPU (markers), energy consumption (bars) and ratio relative to CPU (markers), and average power dissipation (bars) along with its relative ratio to the CPU (markers) for the STFT kernel.]{
        \includegraphics[width=0.7\textwidth]{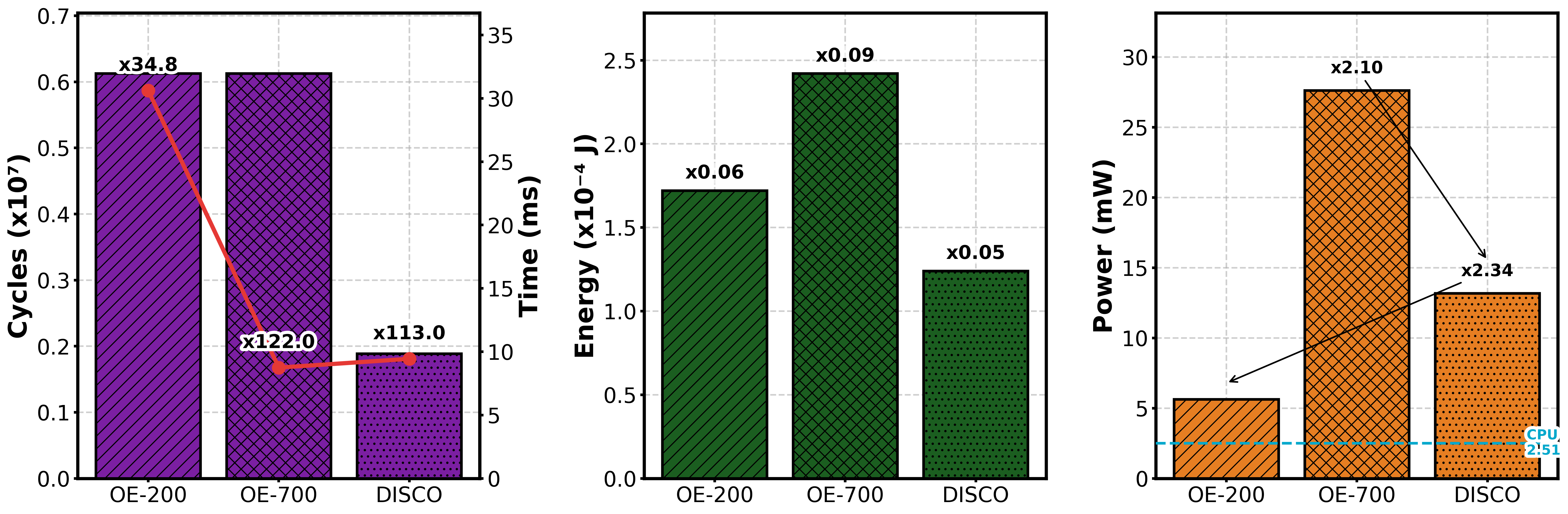}
        \label{fig:transformer_stft}
    }

    % ---  Full Application ---
    \subfloat[End-to-end application results including performance, energy consumption, and power dissipation metrics, incorporating optimized CPU execution for non-offloadable components.]{
        \includegraphics[width=0.7\textwidth]{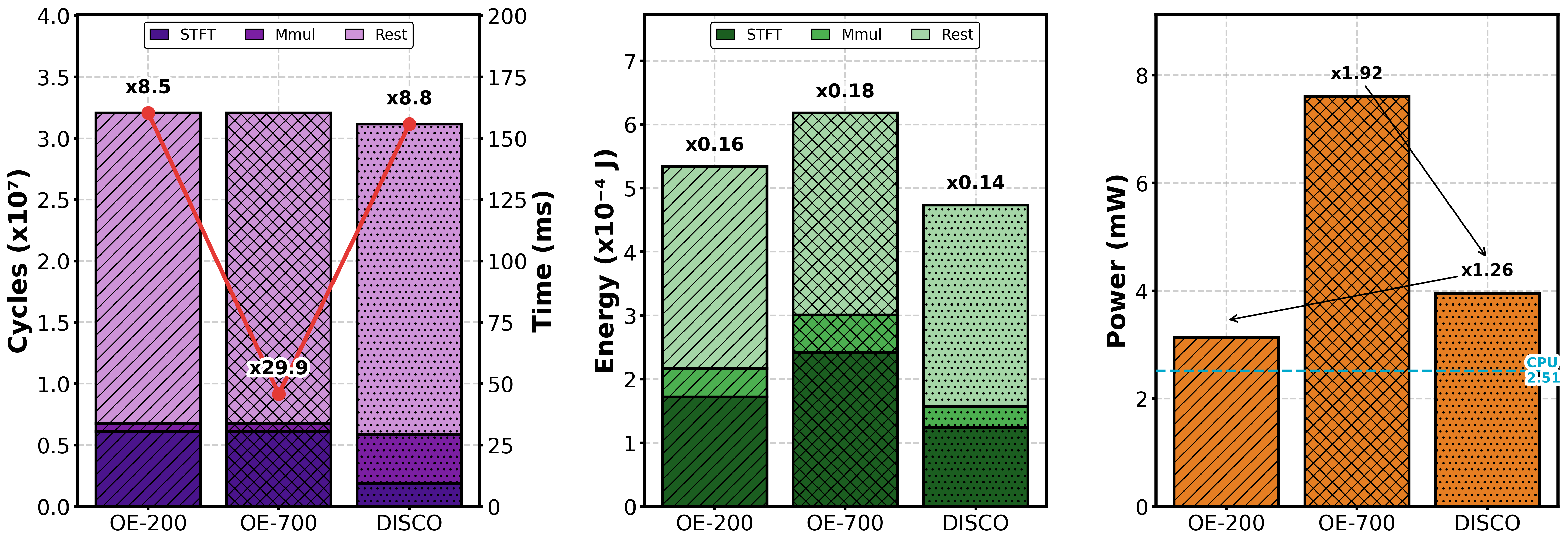}
        \label{fig:transformer_complete}
    }
    
    \caption{Comparative analysis of performance, energy consumption, and power dissipation for the individual STFT kernel and the full end-to-end seizure detection pipeline. For all bar charts, lower values indicate better efficiency or performance. For graphs utilizing markers, larger values indicate better performance for time-related metrics, whereas lower values indicate superior efficiency for energy consumption and power dissipation.}
    \label{fig:transformer_results_fig}
\end{figure*}

Focusing on the STFT algorithm (utilizing the FFT kernel implementation from~\cite{Denkinger2023}) evaluated in Figure~\ref{fig:transformer_stft}, the evaluated architectures exhibit distinct performance dynamics. In terms of execution time, OE-700 achieves the highest performance with a $122\times$ speedup over the optimized CPU baseline, followed closely by DISCO at $113\times$, while OE-200 provides a $34.8\times$ speedup. Power dissipation metrics reveal distinct operating domains for each configuration, following the trend of $\text{OE-200} < \text{DISCO} < \text{OE-700}$. Specifically, OE-200 operates at a modest $5.63\text{~mW}$, DISCO dissipates $13.18\text{~mW}$, and OE-700 peaks significantly higher at $27.62\text{~mW}$. On the energy side, all CGRA configurations demonstrate a massive efficiency advantage over the host processor, with total consumption strictly remaining below $2.5 \times 10^{-4}\text{~J}$. Within the CGRA space, DISCO emerges as the most energy-efficient variant, requiring only $0.05\times$ the energy of the baseline CPU. Conversely, due to its elevated power envelope, OE-700 is the least energy-efficient configuration, though it still limits energy consumption to just $0.09\times$ of the CPU baseline.

Full-application results (Figure~\ref{fig:transformer_complete}) incorporate components unsuitable for CGRA offloading, such as the softmax operation executed on the host CPU. Despite these CPU-bound sections, all CGRA configurations significantly outperform the baseline CPU, which requires $1.37\text{~s}$ of execution time and consumes $3.96 \times 10^{-3}\text{~J}$. Interestingly, at the architectural level, both OE and DISCO exhibit nearly identical clock cycle counts for the full workload. This behavior stems from opposing kernel efficiencies: DISCO demonstrates higher efficiency during the STFT phase, whereas OE performs better during matrix multiplications, resulting in a balanced overall cycle count when both kernels are aggregated. Consequently, when operating at the same clock frequency, OE-200 and DISCO achieve highly comparable performance, delivering speedups of $8.5\times$ and $8.8\times$ over the CPU baseline, respectively. Leveraging its frequency advantage, OE-700 yields the highest performance, accelerating the application by $29.9\times$ relative to the CPU. On the energy side, all CGRA variants tightly constrain consumption between $4.74 \times 10^{-4}\text{~J}$ and $6.18 \times 10^{-4}\text{~J}$. Within this narrow window, DISCO emerges as the most energy-efficient configuration, demanding only $0.14\times$ the energy of the CPU, followed closely by OE-200 at $0.16\times$ and OE-700 at $0.18\times$. In terms of power dissipation, the configurations follow the expected hierarchy ($\text{OE-200} < \text{DISCO} < \text{OE-700}$), with OE-200 dissipating $3.1\text{~mW}$, DISCO requiring $4.0\text{~mW}$, and OE-700 peaking at $7.6\text{~mW}$.

In conclusion, while a heterogeneous domain-specific accelerator utilizing a localized SPM dramatically improves efficiency for intensive data-shuffling workloads like the STFT—matching or exceeding the throughput of high-frequency homogeneous alternatives—the overall end-to-end application energy footprint remains comparable when evaluated at a baseline operating frequency of 200~MHz. The architectural selection therefore follows a strict trade-off hierarchy driven by structural complexity. Baseline homogeneous configurations minimize power dissipation at the expense of prolonged execution cycles. Conversely, operating homogeneous architectures at elevated frequencies minimizes execution time but incurs a steep penalty in peak power dissipation. Integrating structural heterogeneity paired with dedicated memory hierarchies provides a compelling middle ground, successfully balancing localized processing performance against global energy consumption and power dissipation.

% ------------------------------------------------
%                V.C SoA CGRAs
% ------------------------------------------------
\subsection{SoA CGRA Comparison}
\label{sec:sota_comp}

We evaluated OE and DISCO against some existing CGRAs, each synthesized using different technologies, that perform ultra-low power biomedical algorithm execution. To align with the SoA, only the kernel computation cycles of a 256-point FFT are taken into account \cite{de_bruin_r-blocks_2024}. The results are shown in Table \ref{tab:sota_comp_fft}. The maximum frequency of DISCO is between 2-4$\times$ faster than SoA CGRAs. Between the higher frequency and lower number of cycles required to compute the FFT, DISCO performs the computation 3.42$\times$ faster than R-Blocks \cite{de_bruin_r-blocks_2024}, and 37.2$\times$ faster than Riptide \cite{gobieski_riptide_2022}, the latter of which is also implemented in a FinFET technology node. The low execution time results from the very wide data interface between the VWRs and SPM, enabling high-bandwidth parallel data loading, as well as the Shuffle Unit, which performs data shuffling in the exact ``butterfly" data flow pattern of the Cooley-Turkey Radix-2 FFT algorithm in a single clock cycle \cite{cooley_algorithm_1965}.

\begin{table*}[t]
\caption{SoA Post-Synthesis CGRA Comparison on FFT Benchmarks (Adapted from \cite{de_bruin_r-blocks_2024})}
\centering
\begin{tabular}{|l|cc|c|c|c|c|c|}
\hline
 & \multicolumn{2}{c|}{\textbf{OE \cite{alvarez2023open}}} & \textbf{DISCO} & \textbf{VWR2A \cite{denkinger2022vwr2a}} & \textbf{R-Blocks \cite{de_bruin_r-blocks_2024}} & \textbf{Riptide \cite{gobieski_riptide_2022}} & \textbf{SNAFU \cite{gobieski_snafu_2021}} \\ \hline
Technology & \multicolumn{2}{c|}{16nm FinFET} & 16nm FinFET & 40nm CMOS & 22nm FD-SOI & 22nm FinFET & 22nm FinFET \\ \hline
Frequency ($MHz$) & 200 & \textbf{700} & 200 & 80 & 100 & 50 & 50 \\ \hline
Area ($mm^2$) & \textbf{0.061} & 0.088 & 0.30 & N/A & 0.49 & 0.5 & 0.27 \\ \hline
Execution time ($\mu s$) & 116.4 & 33.26 & \textbf{8.80} & 55.45 & 30.08 & 326.92 & 283.78 \\ \hline
Power ($mW$) & 1.62 & 9.47 & 12.4 & 5.41 & 3.99 & \textbf{0.52} & 0.74 \\ \hline
Energy ($\mu J$) & 0.19 & 0.32 & \textbf{0.11} & 0.3 & 0.12 & 0.17 & 0.21 \\ \hline
\end{tabular}
\label{tab:sota_comp_fft}
\end{table*}

On the other hand, OE computes the FFT 3.78$\times$ slower than DISCO even when operating at its maximum frequency due to the lack of internal SPM, which necessitates repeated data fetching from the host processor's main memory. However, its strengths lie in its small area footprint and high maximum frequency. OE is by far the smallest synthesized CGRA among the compared works, requiring between 4.4 and 8.2 times less area than its competitors. This is due to its simple PEs and the fact that its only on-board memory macro is a 2~KiB SRAM for the IMEM. In contrast, DISCO utilizes 42 KiB of SRAM for its heterogeneous instruction memory and wide SPM. This simplicity also enables a high operating frequency of 700 MHz, which is 3.5$\times$ to 14$\times$ faster than SoA CGRAs. Although OE exhibits the second-highest power consumption at its maximum frequency, it should be noted that dynamic power correlates linearly with frequency. If synthesized at 50 MHz (matching Riptide), OE would consume only 0.41 mW—21\% less than Riptide, which currently holds the lowest power consumption among the compared SoA CGRAs.

While DISCO has the highest peak power consumption, its fast execution time results in the lowest energy consumption per FFT. Specifically, DISCO consumes 8.3\%, 35.3\%, and 47.6\% less energy than R-blocks, Riptide, and SNAFU, respectively. The comparison between OE and DISCO reveals a key trade-off between power, energy, and area efficiency. Due to its minimal footprint, OE could be synthesized at 200 MHz for embedded applications requiring miniaturized, ultra-low-power hardware, such as implantable bioelectronics. Conversely, DISCO is more performant and energy-efficient, making it ideal for systems where battery longevity is critical, such as wearable sensors for 24-hour patient monitoring. Furthermore, in High-Performance Computing (HPC) environments, OE could be synthesized at 700 MHz with five parallel replicas. This configuration would roughly occupy the same area as a single DISCO but achieve an ideal execution time of 6.65 $\mu s$—effectively outperforming DISCO by providing higher acceleration within the same area, albeit at the cost of higher energy consumption.

% ------------------------------------------------
%                VI. CONCLUSIONS
% ------------------------------------------------
\section{Conclusions}
\label{sec:conclusions}
This research concludes that while both architectures offer significant improvements over traditional CPUs, their effectiveness depends on specific power, area, and latency requirements.

Regarding the heterogeneous vs. homogeneous performance trade-off, the study finds that the DISCO architecture, equipped with specialized units, is significantly more efficient for data-parallel and data-shuffling kernels like the STFT, achieving the lowest energy consumption per FFT among SoA CGRAs. In contrast, the homogeneous OE architecture is simpler and occupies much less area---roughly $4.4-8.2\times$ less than competitors. This simplicity allows it to scale to a much higher frequency of 700~MHz, where it can match or even exceed the execution performance of more complex architectures like DISCO in several matrix multiplication workloads.

In terms of the impact of local memory on data movement, the inclusion of an SPM drastically reduces memory traffic by exploiting temporal locality, data reuse, and a double-buffering mechanism that masks data transfers.  Memory loads in a memory-less architecture are $8\times$ higher than in the SPM-enabled configuration for representative matrix multiplication kernels, as the memory-less design must repeatedly fetch data from main memory due to its lack of local storage, leading to increased I/O activity and execution stall cycles.

To guide future hardware deployment, this work establishes a clear selection hierarchy governed by resource constraints and environmental envelopes. For ultra-low-power edge nodes with rigid power constraints and tight spatial limits, such as implantable bioelectronics, a low-frequency homogeneous, memory-less CGRA is ideal to minimize standby and operational power consumption at the expense of prolonged execution cycles. In scenarios where processing throughput is critical and the system's power budget is flexible, such as High-Performance Computing (HPC) environments using parallel architectural replicas, a high-frequency homogeneous configuration leverages raw frequency scaling to maximize performance within a compact footprint. Finally, a heterogeneous architecture equipped with local memory represents the optimal choice for energy-constrained systems requiring sustained autonomy, such as wearable sensors for 24-hour patient monitoring, balancing high-throughput execution with global energy efficiency for data-intensive pipelines.

In summary, the choice between these architectures is a trade-off: the heterogeneous approach leverages structural heterogeneity and local storage to maximize data reuse and energy efficiency, while the homogeneous approach offers a streamlined, area-efficient design that utilizes high frequency to achieve competitive performance.

% ------------------------------------------------
%                VII. ACKNOWLEDGEMENTS
% ------------------------------------------------
\section*{Acknowledgements}
The authors acknowledge the use of generative AI tools (Google Gemini and OpenAI ChatGPT) in the preparation of this manuscript. Specifically, these tools were used to refine the clarity, phrasing, and readability of the text across all sections, as well as to generate the Python programming scripts used to plot the pie charts presented in Section~\ref{sec:transformer} and the bar charts presented in Section~\ref{sec:results}. All final text, code, and experimental results were rigorously reviewed and verified by the authors, who accept full accountability for the content.

\bibliographystyle{IEEEtran}
\bibliography{bibtex}

\end{document}